\documentclass[oldversion]{aa} 
\usepackage{graphicx}
\usepackage{longtable}
\usepackage{txfonts}
\usepackage{rotating}
\usepackage{lscape}
\newcommand{\gsim}{\;\lower.6ex\hbox{$\sim$}\kern-7.75pt\raise.65ex\hbox{$>$}\;}
\newcommand{\lsim}{\;\lower.6ex\hbox{$\sim$}\kern-7.75pt\raise.65ex\hbox{$<$}\;}

\begin{document}
\title{NGC~362: another globular cluster with a split red giant 
branch\thanks{Based on observations collected at 
ESO telescopes under programme 083.D-0208}\fnmsep\thanks{
   Tables 2, 3, 4, 5, 6, 7 and 8 are only available in electronic form at the CDS via anonymous
   ftp to {\tt cdsarc.u-strasbg.fr} (130.79.128.5) or via
   {\tt http://cdsweb.u-strasbg.fr/cgi-bin/qcat?J/A+A/???/???}}
 }

\author{
E. Carretta\inst{1},
A. Bragaglia\inst{1},
R.G. Gratton\inst{2},
S. Lucatello\inst{2},
V. D'Orazi\inst{3,4},
M. Bellazzini\inst{1},
G. Catanzaro\inst{5},
F. Leone\inst{6}
Y. Momany\inst{2,7},
\and
A. Sollima\inst{1}
}

\authorrunning{E. Carretta et al.}
\titlerunning{Multiple stellar populations in NGC~362}

\offprints{E. Carretta, eugenio.carretta@oabo.inaf.it}

\institute{
INAF-Osservatorio Astronomico di Bologna, Via Ranzani 1, I-40127 Bologna, Italy
\and
INAF-Osservatorio Astronomico di Padova, Vicolo dell'Osservatorio 5, I-35122
 Padova, Italy
\and
Dept. of Physics and Astronomy, Macquarie University, Sydney, NSW, 2109 Australia 
\and
Monash Centre for Astrophysics, Monash University, School of Mathematical 
Sciences, Building 28, Clayton VIC 3800, Melbourne, Australia
\and
INAF-Osservatorio Astrofisico di Catania, Via S.Sofia 78, I-95123 Catania, Italy
\and
Dipartimento di Fisica e Astronomia, Universit\`a di Catania, Via S.Sofia 78, 
 I-95123 Catania, Italy 
\and
European Southern Observatory, Alonso de Cordova 3107, Vitacura, Santiago, Chile
  }

\date{}

\abstract{We obtained FLAMES GIRAFFE+UVES spectra for both first and 
second-generation  red giant branch (RGB) stars in the globular cluster (GC)
NGC~362 and used them to derive abundances of 21 atomic species for a sample of
92 stars. The surveyed elements include proton-capture (O, Na, Mg, Al, Si),
$\alpha-$capture (Ca, Ti),  Fe-peak (Sc, V, Mn, Co, Ni, Cu), and neutron-capture
elements (Y, Zr, Ba, La, Ce,  Nd, Eu, Dy). The analysis is fully consistent with
that presented for twenty GCs in previous papers of this series.
Stars in NGC~362 seem to be clustered into two discrete groups along the Na-O
anti-correlation, with a gap at [O/Na]$\sim 0$ dex. 
Na-rich, second generation stars show a trend to be
more centrally concentrated, although the level of confidence is not very high.
When compared to the classical second-parameter twin NGC~288, with similar
metallicity, but different horizontal branch type and much lower total mass,
the proton-capture processing in stars of NGC~362 seems to be more extreme,
confirming previous analysis. We discovered the presence of a secondary RGB
sequence, redder than the bulk of the RGB: a preliminary estimate shows that
this sequence comprises about 6\% of RGB stars. Our spectroscopic data and
literature photometry indicate that this sequence is populated almost
exclusively by giants rich in Ba, and probably rich in all s-process
elements, as found in other clusters. In this regards,
NGC~362 joins previously studied GCs like NGC~1851, NGC~6656 (M~22), and
NGC~7089 (M~2).}

\maketitle

\section{Introduction}

In the past few years the paradigm of multiple stellar populations as basic
ingredient of Galactic globular clusters (GCs) has become well assessed (see the
review by Gratton, Sneden \& Carretta 2004, and the recent updates by Martell
2011 and Gratton, Carretta and Bragaglia 2012, including references also to the
photometric evidence). At least two bursts of star formation must have occurred in
GCs, the second generation being formed from a mix of pristine gas and matter
enriched by nuclear processing in the massive stars of the first generation (see
Gratton et al. 2001). The impact of this chain of events is not limited to
the history of GCs, but it probably had a key r\^ole in the building of a
considerable fraction of the Galactic halo (Carretta et al. 2010a, Vesperini et
al. 2010, Martell et al. 2011).
Some important links to global cluster parameters (e.g. total mass, age,
location in the Galaxy, Carretta et al. 2010a) were discovered, and the ouput of
extensive surveys performed with multi-objects facilities like FLAMES (see
Carretta et al. 2006, 2009a,b,c) pointed out that the characteristics of the
nuclear processing occurring in early phases differ from cluster to cluster.
This evidence calls for an in-depth investigation of a number of GCs with
different parameters, like metallicity, concentration, horizontal branch (HB)
morphology. This was the original motivation of our ongoing FLAMES survey
(see Carretta et al. 2006).

In this paper we present an extensive analysis of chemical abundances of
multiple stellar populations in the GC NGC~362. This is a moderately metal-rich 
object ([Fe/H]$=-1.26$~dex\footnote{We adopt the usual spectroscopic notation, 
$i.e.$ [X]= log(X)$_{\rm star} -$ log(X)$_\odot$ for any abundance quantity X,
and  log $\epsilon$(X) = log (N$_{\rm X}$/N$_{\rm H}$) + 12.0 for absolute
number density abundances.} in the 2011 update of the Harris 1996, H96
hereinafter, catalogue  that is based on the metallicity scale of Carretta et al.
2009c) seen projected toward the  Small Magellanic Cloud. 

NGC~362 is variously classified as a red HB (RHB), a young halo (Mackey and van
den Bergh 2005) or an inner halo (Carretta et al. 2010a) cluster. It belongs to
a group of GCs with low total orbital energy (small orbit sizes) and unusual
orbital parameters. The orbit has a chaotic behaviour and is confined close to
the Galactic plane ($z_{max}=2.1$ kpc), with high eccentricity and small
inclination off the orbital plane (Dinescu et al. 1999). An extensive
variability survey in NGC~362 was made by Sz\'ekely et al.  (2007), who find
more than 45 RR Lyrae variables, a metallicity [Fe/H]$=-1.16$~dex, and a mean
period of RRab stars of $0.585\pm0.081$~days, which places this cluster in the
Oosterhoff type I group.

The pair NGC~362-NGC~288 is considered a classical example  of second parameter
clusters: the colour of the HB of these two GCs is very different (NGC~362: red;
NGC~288: blue) despite their similar metal-abundance.  
Catelan et al. (2001) compared the relative age provided by the HB morphology
with that obtained from the main sequence for the pair NGC~362-NGC~288. They
used the bimodal HB of NGC~1851 as a ``bridge" (see also Bellazzini et al.
2001). They found that NGC~362 is about 2 Gyr  younger than NGC~288, supporting
the concept that age is a main second parameter (beside metallicity) in shaping
the distribution of stars along the HB. The same result was obtained by Dotter
et al. (2010) and Gratton et al. (2010) from the analysis of homogeneous
photometric and spectroscopic databases. However, while Catelan et al. claimed
that the mass  dispersion on the HB of NGC~362 is substantially larger than for
NGC~288, the opposite was found by Gratton et al. According to their analysis,
the mass spread required to account for the HB morphology in NGC~362 is only
0.004 $M_\odot$ against 0.024 $M_\odot$ for NGC~288. This is due to the adoption
of an universal mass loss law with a linear dependence on metallicity, that well
reproduces the median colours of HB stars. 

Piotto et al. (2012) found that a few (actually, less then 3\%)
subgiant branch stars of NGC~362 are fainter than the other subgiants
of similar colour. This faint sequence might be interpreted as a small group
of objects having either older ages or larger total CNO content than the
majority of the stars.

Abundance analyses of stars in NGC~362 are rather scanty.  Several studies based
on low dispersion spectroscopy determined the pattern of the CN and CH
distribution in NGC~362, typically found to be bimodal (Smith 1983, 1984; Kayser
et al. 2008, Smith and Langland-Shula 2009 and reference therein). Shetrone and
Keane (2000) derived abundances for several elements in a dozen of giants  from
high  dispersion spectra and compared their chemical pattern to that derived
from 13  giants in NGC~288. At the epoch (before the analysis of O, Na, Mg, Al
in  unevolved cluster stars by Gratton et al. 2001) it was still debated how
much evolutionary (mixing) effects were superimposed to  {\it ab initio}
abundance variations. However, even from their moderate-size samples, Shetrone
and Keane (2000) were able  to point out differences in zero-point and slope
existing among the observed Na-O  anti-correlation in this pair of GCs. Recently
Worley and Cottrell (2010) used high dispersion spectroscopy to study the
pattern of light and heavy neutron-capture process elements in NGC~362.

In our study we enlarge the sample of stars analysed with moderate-high
resolution spectroscopy to more than 90 red giant branch (RGB) stars in NGC~362,
obtaining homogeneous abundances of proton-capture, $\alpha-$capture, Fe-peak
and neutron-capture elements.

The paper is organized as follows: observations are presented in Section 2; radial
velocities and a brief discussion of the kinematics of the cluster are in 
Section 3; Section 4 presents the derivation of the atmospheric parameters and
the abundance analysis; results of this analysis are given in Section 5; finally,
discussion and conclusions are in Section 6. The Appendix presents the derivation
of errors in the abundance analysis.

\begin{figure}
\centering
\includegraphics[scale=0.44]{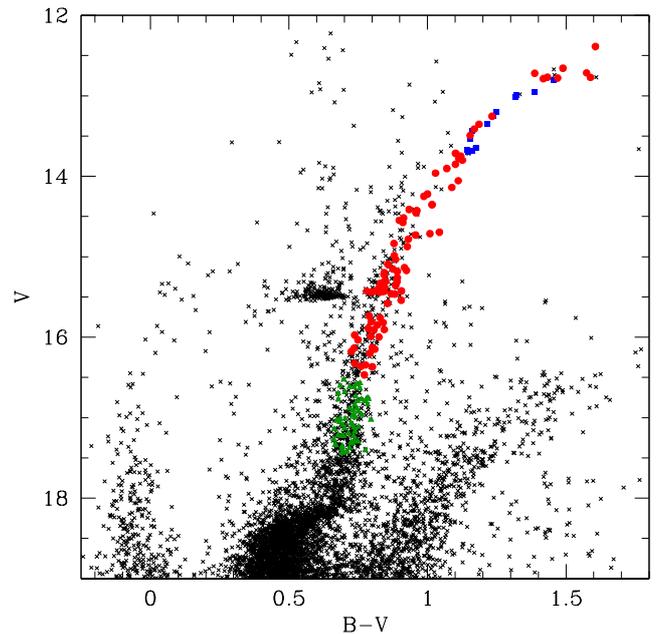}
\caption{The $V,B-V$ CMD of NGC~362 (light grey crosses). Stars selected for the
present study are plotted as filled, larger symbols: blue squares are the stars 
observed with UVES, red circles are stars with GIRAFFE spectra, and green
triangles are stars observed with GIRAFFE but not analyzed (see text). The faint
blue and red sequences are due to the young main sequence stars and old red
giants of the Small Magellanic Cloud.}
\label{f:cmdsel362}
\end{figure}

\section{Observations}

Our targets were selected from unpublished Johnson $B,V$ Wide Field Imager (WFI)
photometry obtained at the 2.2m ESO-Max Planck Telescope (La Silla, Chile) and
astrometrised by one of us (Y. Momany), integrated with $K$ band magnitudes 
from the Point Source Catalogue of 2MASS (Skrutskie et al. 2006).  The $V,B-V$
colour magnitude diagram (CMD) of NGC~362 is shown in Fig.~\ref{f:cmdsel362}
with superimposed stars of our spectroscopic sample, chosen to be near the RGB
ridge line and with no close companion. NGC~362 is seen projected against the
Small Magellanic Cloud, which is responsible for the faint blue and red
sequences visible in that diagram; they are young main sequence stars and old
red giants, respectively.

The log of the observations is given in Table~\ref{t:logobs}; we obtained two
exposures with the HR11 high resolution grating covering the Na~{\sc i}
5682-88~\AA\ doublet and two exposures with the grating HR13 including the
[O~{\sc i}] forbidden lines at 6300-63~\AA.  We observed a total of 14 (bright)
giants with the fibres feeding the UVES spectrograph (Red Arm, with spectral
range from 4800 to 6800~\AA\ and R=45,000;  blue squares in
Fig.~\ref{f:cmdsel362}) and 136 RGB stars with GIRAFFE (filled circles and
triangles). All stars turned out to be (likely) member of the cluster based on radial
velocities (RV). However, for the present analysis we retained only stars with
effective temperature below 5200 K because for warmer stars (green triangles) 
the low S/N and noisy spectra prevented us from measuring accurate enough
equivalent widths ($EW$s)\footnote{The hottest star left in our final sample is
\#5447, with an effective temperature of 5203 K.}. This limit corresponds to a
magnitude level $V=16.47$, indicated by a solid line in the CMD.

We used the 1-D, wavelength calibrated spectra as reduced by the ESO personnel 
with the dedicated FLAMES pipeline. Radial velocities (RV) for stars observed with
the GIRAFFE spectrograph were obtained using the 
{\sc IRAF}\footnote{IRAF is  distributed by the National Optical Astronomical
Observatory, which are operated by the Association of Universities for Research
in Astronomy, under contract with the National Science Foundation } task 
{\sc FXCORR}, with appropriate templates, while those of the stars observed
with UVES were derived with the IRAF task RVIDLINES.

\begin{table}
\centering
\caption{Log of FLAMES observations for NGC 362}
\setlength{\tabcolsep}{1.5mm}
\begin{tabular}{cccccc}
\hline\hline
   Date         &     UT       & exp. & grating & seeing & airmass\\
                &              & (sec)&         &  (")   &        \\ 
\hline
 Aug. 04, 2009  &08:07:18.230  &  2600   & HR11    & 1.02   & 1.454  \\
 Aug. 04, 2009  &08:57:24.456  &  2600   & HR11    & 0.88   & 1.442  \\
 Aug. 05, 2009  &07:27:25.601  &  2600   & HR13    & 0.65   & 1.482  \\
 Aug. 05, 2009  &08:12:33.313  &  2600   & HR13    & 0.86   & 1.449  \\
\hline
\end{tabular}
\label{t:logobs}
\end{table}

One star (13875) observed with GIRAFFE shows the TiO band head at 6158~\AA\ in
the HR13 spectrum (see Valenti et al. 1998). This star has many peculiarities
beside being of spectral type M: it is a variable (V2: Clement 1997) with a 
period of 90 days; it was found to be very Li-rich by Smith et al. (1999); and 
it presents carbon dust according to Boyer et al. (2009). It was discarded from the 
following analysis.
 
We have 12 stars in common between the UVES and GIRAFFE datasets; disregarding
star V2 our final sample in NGC~362 consists of 92 RGB stars. Coordinates,
magnitudes and heliocentric RVs are shown in Tab.~\ref{t:coo36} (the full table
is only available in electronic form at CDS). 

\section{Kinematics}

As discussed in Bellazzini et al.~(2012; B12 hereinafter) the samples of RV
estimates that are obtained as a natural by product of this project are not
ideal for kinematical analyses. However, because of the fibre allocation
constraints described in B12, in many cases they are nicely complementary to
existing dataset as they preferentially probe the outskirts of the clusters.
This allowed us to reveal rotation signals in the outer regions of several
clusters that went unnoticed in previous studies.  In the following - as for the
chemical analysis - we use only stars cooler than 5200~K (red circles in Fig.~1
and below) since only for these stars the membership is fully ascertained based
on both RV and chemical composition, following B12. Stars warmer than this
values are plotted as green circles (as in Fig.~1) in the figures presenting
kinematic results, for completeness. We perform the  same kind of analysis as in
B12, using the same techniques: we refer to that paper for details and
discussion.

The detailed analysis by Fischer et al. (1993: F93) is based on 210 RV members
of this cluster, all of them lying within $R=4.0\arcmin$. On the other hand, our
sample cover the range  $1.0\arcmin\le R\le 11.0\arcmin$, reaching the tidal
radius of the clusters ($r_t=10.3\arcmin$, H96). We merged our catalog with the
one by F93, keeping our RV when estimates from both sources are available. From
a subset of 38 stars in common (excluding two stars classified as binary from
RVs at two epochs by F93, our stars 11413 and 18947\footnote{All the stars
classified as binaries by F93 have been excluded from the analysis of the
cluster kinematics. The four excluded stars are H1348, H1419, H2205, H2222, in
the nomenclature of F93. H1419 = 18947 and H2205= 11413 are also included in our
own sample and in the following chemical analysis.}) we find a
mean difference of $\Delta$RV=-1.15 km s$^{-1}$ with $\sigma=1.54$; this small
zero point offset was corrected before merging the two samples. The errors on
individual RV estimates from the two datasets are very similar ($\simeq 1.0$ km
s$^{-1}$).

\begin{figure}
\centering
\includegraphics[scale=0.47]{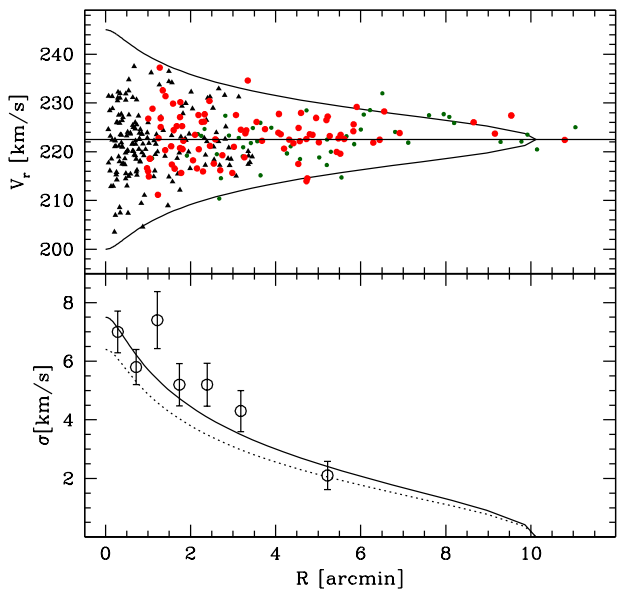}
\caption{Upper panel: radial velocity as a function of the distance from the
cluster centre. Colored symbols are the same as in Fig.~1; filled triangles are
stars from F93.  The solid lines represents the $\pm 3\sigma$ envelope of the
King (1966) model that best fits the surface brightness profile, according to
H96. The profile is normalized to the central velocity dispersion
$\sigma_0=7.5$~km s$^{-1}$.  Lower panel: velocity dispersion profile as derived
from the global sample (this work + F93, excluding stars lacking abundance
estimates). The solid line is the profile of the same King (1966) model as
above, the dotted line is the same model normalized at the central velocity
dispersion reported in H96, $\sigma_0=6.4$~km s$^{-1}$.}
\label{f:RVrC}
\end{figure}

In the lower panel of Fig.~\ref{f:RVrC} we show that the velocity dispersion
profile we obtain from our global sample is hardly compatible with a King (1966)
model having the best-fit parameters reported in H96 and scaled to the central
velocity dispersion reported there for this cluster, i.e.  $\sigma_0=6.4$~km
s$^{-1}$. In the upper panel it can be appreciated that even adopting a higher
value, providing an acceptable fit to the observed profile ($\sigma_0=7.5$~km
s$^{-1}$), there are obvious members lying outside the $\pm 3-\sigma$ envelope
of the model, suggesting that some unbound or partially-bound extra-tidal stars
are present, or that King models are not fully adequate to describe the
structure and dynamics of the system (see McLaughlin \& van der Marel 2005, and
Correnti et al. 2011 for discussion and references).

\begin{figure}
\centering
\includegraphics[scale=0.44]{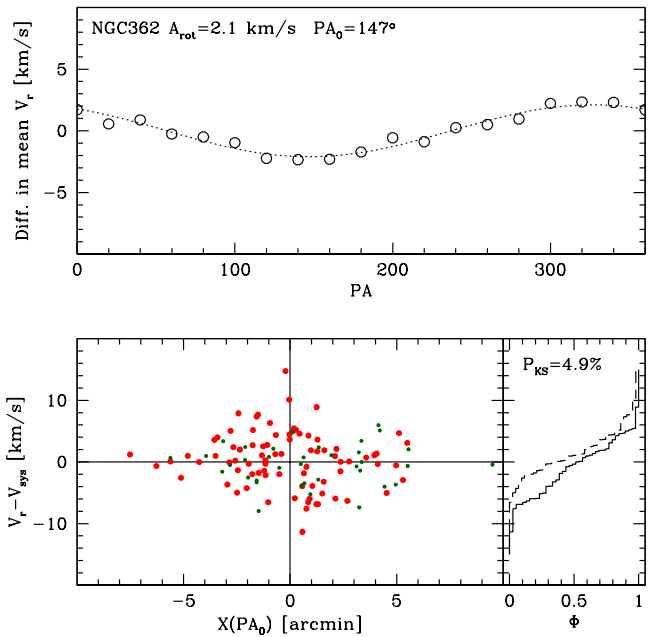}
\caption{Rotation in NGC~362. Upper panel: difference between the mean
velocities on each side of the cluster with respect to a line passing through
the cluster centre with a position angle (PA, measured from north to east,
north=0$^o$, east=90$^o$), as a function of the adopted PA. Our preferred 
solution is represented by the dotted line and is obtained from bona-fide members
from our sample (i.e. stars with abundance estimates).
Lower panels: the rotation curve of NGC~362.
In the left panel the RV in the system of the cluster is plotted as a function
of distance from the centre projected onto the axis perpendicular to the best
fit rotation axis found in the upper panel. the meaning of the symbols are the same as 
in Fig.~1 and 2, above. The right panel shows the comparison
of the cumulative RV distributions of stars having $X(PA_0)>0.0$ (continuous
line) and $X(PA_0)<0.0$ (dotted line).}
\label{f:rotaz}
\end{figure}

Looking for rotation in the global sample we find no significant signal, in
agreement with F93. However if we limit to our sample  we find the
$A_{rot}=2.1$~km s$^{-1}$, and $PA=147\deg$ signal shown in Fig.~\ref{f:rotaz},
suggesting that  some rotation may be indeed present in the outer regions. Note
that the warmer stars excluded from the analysis (green points) appear to share
the same pattern. The trend toward $V_r-V_{sys}\sim 0.0$ at large absolute
values of $X(PA_0)$ is more consistent with an intrinsic nature of the kinematic
pattern, as a velocity gradient induced by Galactic tides would show larger
amplitudes at larger distances. NGC~362 nicely fits into the correlations
recently found by B12.

Finally, we note that the velocity dispersion changes as a function of the Na
abundance, Na-poor stars having lower dispersion than Na-rich ones. This
is due to the radial trend shown by the different concentration of Na-poor and
Na-rich stars, the first preferentially populating more external regions (see
below).
There is no significant trend of the rotation pattern with Na abundance,
but it is very likely that our sample is too sparse to investigate such subtle
effects.

\section{Atmospheric parameters, abundance analysis, and metallicity}

Only a brief summary of the analysis methods will be reported here, since the
atmospheric parameters were derived following the same procedure adopted 
for the other GCs targeted by our FLAMES survey (see e.g. Carretta et al.
2009a,b). 

First-pass input effective temperature $T_{\rm eff}$\ (from $V-K$ colours) and
bolometric corrections were derived from the calibrations of  Alonso et al.
(1999, 2001). 
These values were refined using a relation between $T_{\rm eff}$\  and $K$
magnitudes\footnote{For a few stars with no $K$ in 2MASS we obtained an estimate
of the $K$ values by interpolating a quadratic relation as a function of $V$ 
magnitudes.}.
Gravities were obtained from stellar masses and radii, these  last derived from
luminosities and temperatures.  The  reddening and the distance modulus for
NGC~362 were taken from the Harris  (1996) catalogue (2011 update). We adopted a
mass of 0.85~M$_\odot$\  for all stars and $M_{\rm bol,\odot} = 4.75$ as the
bolometric magnitude for the Sun, as in our previous studies.

The abundance analysis mainly rests on equivalent widths ($EW$s). We checked 
that $EW$s measured on the GIRAFFE spectra are on the same system defined by 
high-resolution UVES spectra.  The values of the microturbulent velocities
$v_t$\  were obtained by eliminating trends between Fe~{\sc i} abundances and
expected  line strength (Magain 1984). Finally, models with the appropriate
atmospheric parameters and whose abundances matched those derived from 
Fe {\sc i} lines were interpolated  within the Kurucz (1993) grid of model
atmospheres (with the option for overshooting on) to derive the final
abundances. The adopted atmospheric parameters and iron abundances are listed in
Tab.~\ref{t:atmpar36}.

Our procedure for error estimates is detailed in Carretta et al. (2009a,b);
results, with a brief description, are given in the Appendix for UVES and 
GIRAFFE observations (see Tab.~\ref{t:sensitivityu36} and
Tab.~\ref{t:sensitivitym36}, respectively).

Oxygen abundances were obtained from the [O~{\sc i}] line at 6300.3~\AA\ (after 
cleaning this spectral region from telluric lines as described in Carretta et
al. 2007a) and, whenever possible, from the [O~{\sc i}] line at 6363.8~\AA.
Sodium abundances, derived from the $EW$s of the 5682-88 and 6154-60~\AA\
doublets, were corrected for departures from the LTE assumption following Gratton
et al. (1999). Abundances of O, Na, Al (from the 6696-99~\AA\ doublet), and Mg
(derived as in Carretta et al. 2009b) for individual stars are listed in
Tab.~\ref{t:proton36}.

Beside Mg (reported among elements involved in proton-capture processes), we
measured the abundances of the $\alpha-$elements Si, Ca, and Ti either  from
UVES or GIRAFFE spectra (see Tab.~\ref{t:alpha36})\footnote{The case of Si as an
element involved in the p-capture reactions is discussed in Section 5.1}. We
measured Ti abundances from lines of both neutral and singly ionised species on
UVES spectra (with larger spectral coverage). On average, the abundances
obtained from these two ionization states are in excellent agreement with each
other (see below), supporting our adopted scale of atmospheric parameters.

We obtained abundances for the Fe-peak elements Sc~{\sc ii}, V~{\sc i},
Cr~{\sc i}, Co~{\sc i}, and Ni~{\sc i} (stars with GIRAFFE and UVES spectra)
and additionally Mn~{\sc i} and Cu~{\sc i} for stars with UVES spectra only.
The corrections due to the hyperfine structure (see Gratton et al. 2003 for
references) were applied  whenever relevant (e.g. Sc, V, Mn, Co).
We obtained the abundances of several neutron capture elements 
(Y~{\sc ii}, Zr~{\sc ii}, Ba~{\sc ii}, La~{\sc ii}, Ce~{\sc ii}, Nd~{\sc ii}, 
Eu~{\sc ii}, and Dy~{\sc ii}, mostly from UVES spectra) using a 
combination of  spectral synthesis and $EW$ measurements. Details on
transitions can be found in Carretta et al. (2011).

Abundances of Fe-peak elements for individual stars are reported in
Tab.~\ref{t:fegroup36}. Results for neutron capture elements are given in
Tab.~\ref{t:neutron36} and Tab.~\ref{t:zrba36} for stars with UVES and GIRAFFE
spectra, respectively. The averages of all measured elements with their $r.m.s.$
scatter are listed in  Tab.~\ref{t:meanabu36}.

\setcounter{table}{8}
\begin{table}
\centering
\caption{Mean abundances from UVES and GIRAFFE}
\begin{tabular}{lcc}
\hline
                     &               &               \\
Element              & UVES	     & GIRAFFE       \\
                     &n~~   avg~~  $rms$ &n~~	avg~~  $rms$ \\        
\hline
$[$O/Fe$]${\sc i}    &14   +0.14 0.20 &64   +0.89 0.18\\
$[$Na/Fe$]${\sc i}   &14   +0.19 0.19 &90   +0.11 0.25\\
$[$Mg/Fe$]${\sc i}   &14   +0.33 0.04 &84   +0.33 0.04\\
$[$Al/Fe$]${\sc i}   &14   +0.24 0.19 &  	      \\
$[$Si/Fe$]${\sc i}   &14   +0.22 0.04 &87   +0.26 0.04\\
$[$Ca/Fe$]${\sc i}   &14   +0.30 0.03 &89   +0.34 0.02\\
$[$Sc/Fe$]${\sc ii}  &14 $-$0.03 0.05 &90 $-$0.07 0.04\\
$[$Ti/Fe$]${\sc i}   &14   +0.22 0.04 &84   +0.16 0.03\\
$[$Ti/Fe$]${\sc ii}  &14   +0.21 0.05 &  	      \\
$[$V/Fe$]${\sc i}    &14 $-$0.03 0.03 &86 $-$0.05 0.03\\
$[$Cr/Fe$]${\sc i}   &14 $-$0.02 0.07 &88 $-$0.03 0.04\\
$[$Mn/Fe$]${\sc i}   &14 $-$0.33 0.04 &    	      \\
$[$Fe/H$]${\sc i}    &14 $-$1.17 0.05 &90 $-$1.17 0.04\\
$[$Fe/H$]${\sc ii}   &14 $-$1.21 0.08 &69 $-$1.18 0.06\\
$[$Co/Fe$]${\sc i}   &14 $-$0.28 0.06 &36 $-$0.05 0.09\\
$[$Ni/Fe$]${\sc i}   &14 $-$0.13 0.03 &90 $-$0.09 0.04\\
$[$Cu/Fe$]${\sc i}   &14 $-$0.50 0.12 &               \\
$[$Zn/Fe$]${\sc i}   &12   +0.21 0.06 &43   +0.28 0.08\\  
$[$Y/Fe$]${\sc ii}   &14   +0.07 0.11 &  	      \\  
$[$Zr/Fe$]${\sc ii}  &14   +0.50 0.12 &  	      \\  
$[$Ba/Fe$]${\sc ii}  &14   +0.30 0.27 &68   +0.18 0.21\\  
$[$La/Fe$]${\sc ii}  &14   +0.33 0.09 &  	      \\  
$[$Ce/Fe$]${\sc ii}  &14   +0.14 0.12 &  	      \\  
$[$Nd/Fe$]${\sc ii}  &14   +0.35 0.10 &  	      \\  
$[$Eu/Fe$]${\sc ii}  &14   +0.70 0.07 &  	      \\  
$[$Dy/Fe$]${\sc ii}  &14   +0.68 0.13 &  	      \\  
\hline
\end{tabular}
\label{t:meanabu36}
\end{table}

The mean metallicity we found for NGC~362 from stars with UVES spectra is 
[Fe/H]$=-1.168 \pm0.014 \pm0.051$ dex ($rms=0.052$ dex, 14 stars), where the
first error bar is from statistics and the second one refers to the systematic
effects. From the large sample of stars with GIRAFFE spectra we derived a value of 
[Fe/H]$=-1.174 \pm0.004 \pm0.062$ dex ($rms=0.041$ dex, 90 stars).
The metal abundance of NGC~362 is very similar to that of NGC~1851 ([Fe/H]$=-1.19$ dex, 
Carretta et al. 2011) and 0.14 dex higher than the metal content of NGC~288
([Fe/H]$=-1.31$ dex, Carretta et al. 2009c) on our homogeneous metallicity scale.

\begin{figure}
\centering 
\includegraphics[bb=78 171 460 691, clip,scale=0.52]{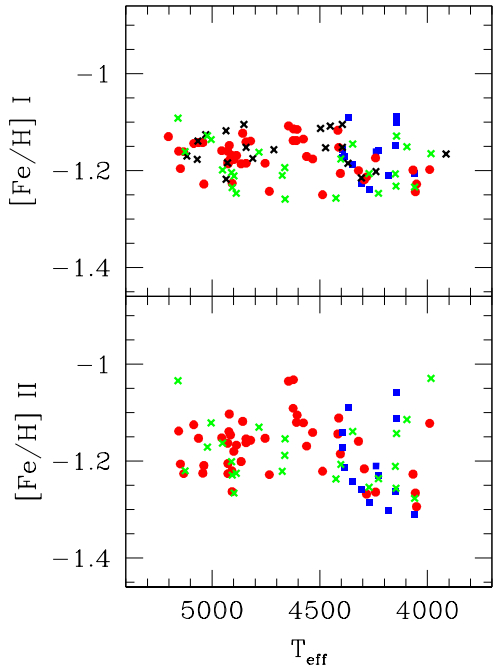}
\caption{Abundance ratios [Fe/H] {\sc i} (upper panel) and [Fe/H] {\sc ii} (lower
panel) as a function of T$_{\rm eff}$ for all analysed stars. Blue
squares are stars with UVES spectra, filled circles are those with GIRAFFE
spectra observed with both HR11 and HR13 gratings, whereas crosses indicate
stars observed with only the HR11 (black) or the HR13 (green) grating.}
\label{f:feteff36}
\end{figure}

The abundances of iron obtained from singly ionized species are in nice
agreement with those from neutral lines: [Fe/H]{\sc ii}$=-1.21$ ($rms=0.08$, dex
14 stars) from UVES and  [Fe/H]{\sc ii}$=-1.18$ ($rms=0.06$ dex, 69 stars) from
GIRAFFE. They do not present any trend as a function of the
effective temperature, as shown in Fig.~\ref{f:feteff36}.

Our Fe abundance is in very good agreement with that listed in the old (1996)
version of the Harris catalogue and with that obtained by Szekely et al. (2007)
from analysis of the pulsational properties of RR Lyrae. It
is higher (although in agreement within the uncertainties) than the
value listed in the most recent version of the Harris catalogue, and of the
average values obtained by Shetrone and Keane (2000) and Kraft and Ivans (2003). 

\section{Results}

\subsection{Chemistry of multiple populations in NGC~362}

\begin{figure}
\centering 
\includegraphics[scale=0.40]{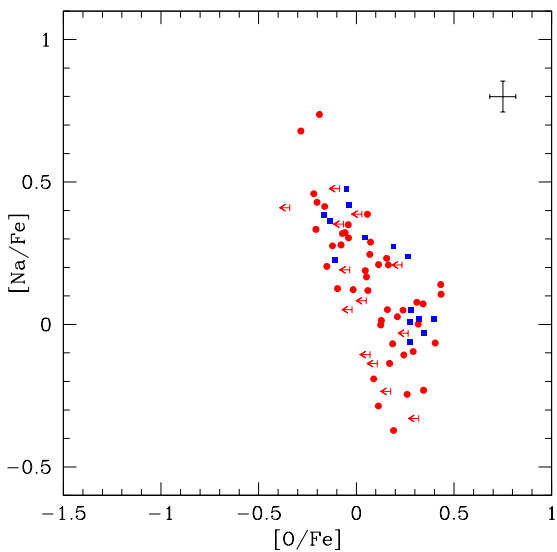}
\caption{The Na-O anti-correlation observed in NGC~362. Blue squares are stars
observed with UVES, while filled circles indicate stars with GIRAFFE spectra.
Upper limits in O are shown as arrows, and star-to-star (internal) error bars
are plotted in the upper-right corner of the plot.}
\label{f:m36antitot}
\end{figure}

The Na-O anti-correlation in NGC~362 is based on 71 stars for which we were able
to provide O and Na abundances from UVES and/or GIRAFFE spectra (we have
58 detections and 13 upper limits for O; the total number of stars with
[Na/Fe] values is 92).
Results are shown in Fig.~\ref{f:m36antitot}.

\begin{figure}
\centering 
\includegraphics[scale=0.40]{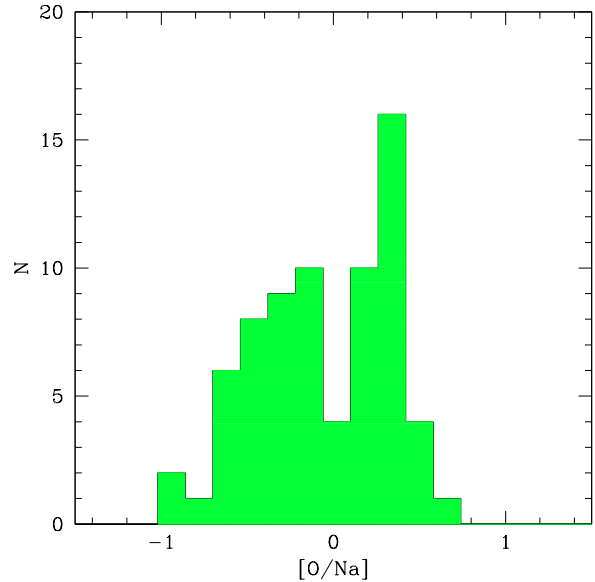}
\caption{The distribution of the [O/Na] ratios in NGC~362 from the combined
sample UVES+Giraffe.}
\label{f:histo}
\end{figure}

The interquartile range IQR for the ratio [O/Na] in this cluster is 0.644 dex. This
value nicely fits into the correlation with the total absolute magnitude  of the
cluster ($M_V=-8.41$, Harris 1996) established in Carretta et al. (2010a).
NGC~362 also participates to the relation with the maximum temperature along  the
HB ($\log T_{\rm eff}=4.079$, from Recio-Blanco et al. 2006), discovered by
Carretta et al. (2007b) and repeatedly updated and strengthened during our
FLAMES survey. NGC~362 stars seem to be clustered along the Na-O
anti-correlation in two distinct groups, one with high O and low Na and the other
more enriched in Na and depleted in O. This sub-division is even more evident if
we  consider only the 14 stars observed at higher resolution with UVES, and it
is nicely supported by the distribution of the [O/Na] ratios shown in
Fig.~\ref{f:histo}, that maximise the signal along the Na-O anti-correlation.
We also note that the same separation into two groups was already present 
in the independent analysis by Shetrone and Keane (2000) (see their Fig.2).

Using the quantitative criteria introduced by Carretta et al. (2009a) we can 
use O and Na abundances to quantify the size of the primordial (P) population
and of the intermediate (I) and extreme (E) components of the second generation
stars. We found that the fractions of P, I, and E stars for NGC~362 are
$22\pm6\%$,  $75\pm10\%$, and $3\pm2\%$, respectively. 
We note that according to these criteria, the lower Na-high-O group 
visible in Fig.~\ref{f:m36antitot} includes not only the P component 
([Na/Fe]$<-0.03$ dex, in NGC~362), but also part of the intermediate fraction I.
We warn however that the distribution of stars along the Na-O anticorrelation
from GIRAFFE spectra may be not optimally suited to study the discrete vs
continuous nature of multiple populations.

\begin{figure}
\centering 
\includegraphics[scale=0.40]{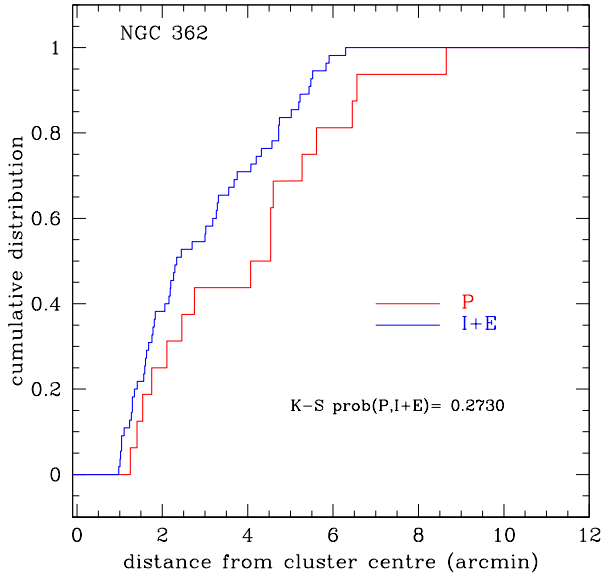}
\caption{Cumulative distribution of the stars of first (primordial P component)
and second generation (the merged I+E components together) as a function of the
projected distance from the cluster centre.}
\label{f:distrPIE}
\end{figure}

The radial distribution of these components is shown in Fig.~\ref{f:distrPIE},
where the second generation stars of the I and E component are merged together,
since only two stars belong to the E fraction. We find that second generation
stars are more centrally concentrated, a result already found in several other
clusters (see e.g. Lardo et al. 2011, Nataf et al. 2011, Milone et al. 2012).

\begin{figure}
\centering 
\includegraphics[scale=0.40]{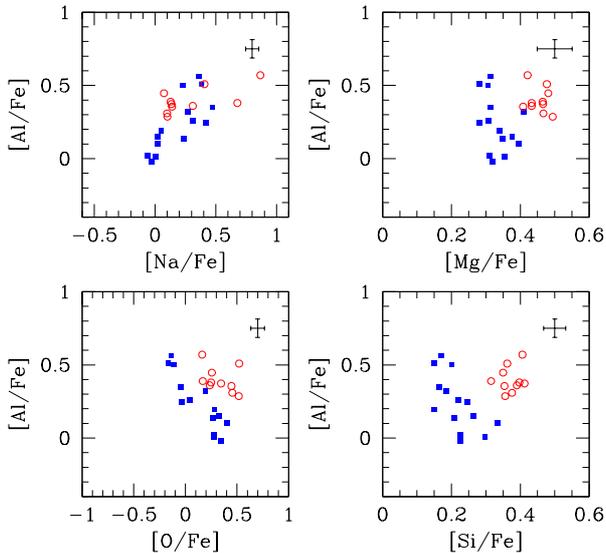}
\caption{Relations of the [Al/Fe] ratios in NGC~362 (filled blue squares) from
UVES spectra as a function of  [Na/Fe] (upper left panel), [Mg/Fe] (upper
right), [O/Fe] (lower left panel), and [Si/Fe] (lower right). The star to star
error  bars are indicated in each panel. As a comparison, the open circles are
results for giants in NGC~288 from Carretta et al. (2009b).}
\label{f:light4u36}
\end{figure}

The interplay of the various elements involved in  proton-capture processes in
NGC~362 is summarized in Fig.~\ref{f:light4u36} for the stars observed with
UVES. For this sub-sample, many light element abundances (O, Na, Mg, Al,
Si) are available from our data, thanks to the large spectral coverage. As a
reference, we also plot in this figure the results for 10 RGB stars observed
with UVES in NGC~288 by Carretta et al. (2009b). The range of Al variations is
not extreme among giants in NGC~362, yet it is almost twice the spread in Al
observed in NGC~288. The Al-O anti-correlation is well developed in NGC~362,
whose O abundances reach a level significantly lower than in NGC~288, as found
by Shetrone and Keane (2000). All these features can be explained
by the larger mass of NGC~362, which is almost five times more massive than
NGC~288 (see McLaughlin and van der Marel 2005).

\begin{figure}
\centering 
\includegraphics[scale=0.40]{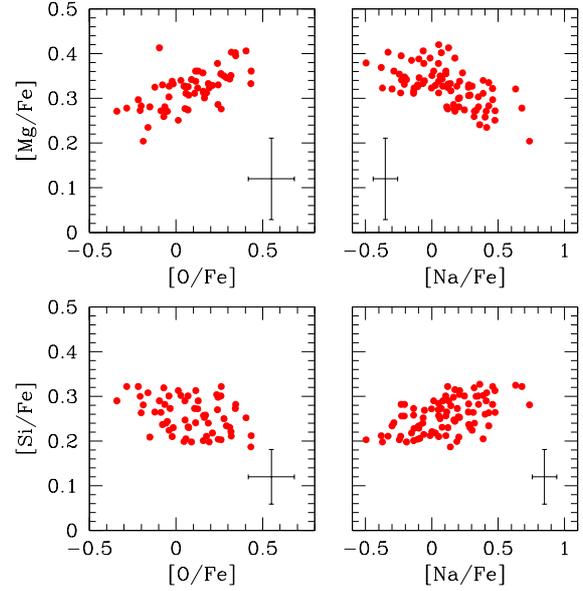}
\caption{[Mg/Fe] ratios (upper panels) as a function of [O/Fe] (left) and
[Na/Fe] (right) ratios from GIRAFFE spectra. The same relations for [Si/Fe] are
plotted in the lower panels. The internal errors are indicated in each panel.}
\label{f:mgsi36}
\end{figure}

Due to the limited size of the sub-sample of stars with UVES spectra, it is not
possible to evince whether the polluting matter producing second generation
stars was processed at very high temperature. As discussed by Yong et al. (2005)
and Carretta et al. (2009b), when the H-burning temperature exceeds 65 MK a
leakage from the Mg-Al cycle on $^{28}$Si occurs and some amount of Si is
produced at the expense of Mg, beside the main outcome, Al. To verify that this
is the case also for NGC~362 we show in Fig.~\ref{f:mgsi36} the run of [Mg/Fe]
and [Si/Fe] ratios as a function of O and Na for the much  larger sample of
giants observed with GIRAFFE in this cluster.

Mg is correlated to elements depleted in proton-capture reactions and
anti-correlated to elements which are enhanced in this burning, and the opposite 
is seen to occur for Si. All these relations are statistically robust, the level
of confidence formally exceeds 99\% in all four cases. However, the relations
involving Si seem to be driven by a few stars only. While the stars with UVES
spectra give Mg abundances that nicely fall on the relations defined by the
GIRAFFE sample involving Mg, the agreement is not so satisfactory concerning Si.
In summary, it is not clear that the first generation polluters were of the
right mass (inner temperature) range to significantly change the Si abundance
above the level typical of enrichment from type II supernovae. 
In NGC~362 we did not observe the typical Si-Al correlation that in massive
or metal-poor GCs (such as NGC~2808 or NGC~6752, Carretta et al. 2009b) suggest
that Si is partly produced by proton-capture reactions, beside the classical
$\alpha-$capture processes.

\subsection{Other elements}

\begin{figure}
\centering 
\includegraphics[scale=0.42]{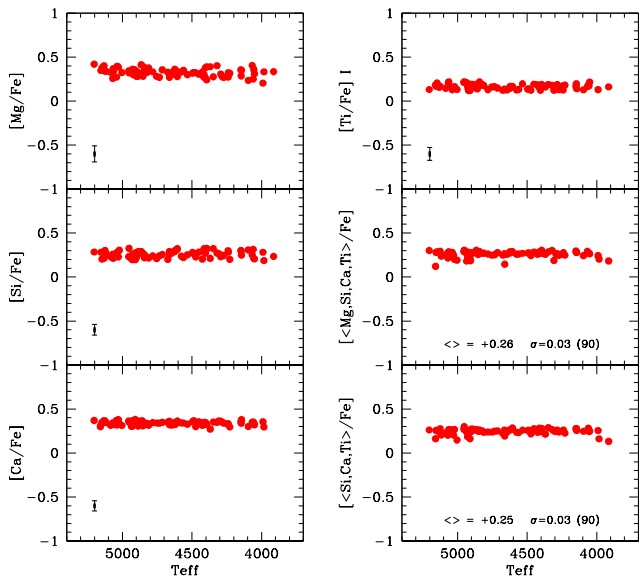}
\caption{Abundance ratios of $\alpha-$elements Mg, Si, Ca, Ti~{\sc i} as a
function of the effective temperature. The average of [$\alpha$/Fe] ratios are
shown  in the last two panels on the right column (including and excluding the
Mg abundance from the mean, respectively). Error bars indicate internal
star-to-star errors.}
\label{f:alpm36}
\end{figure}

\begin{figure}
\centering 
\includegraphics[scale=0.42]{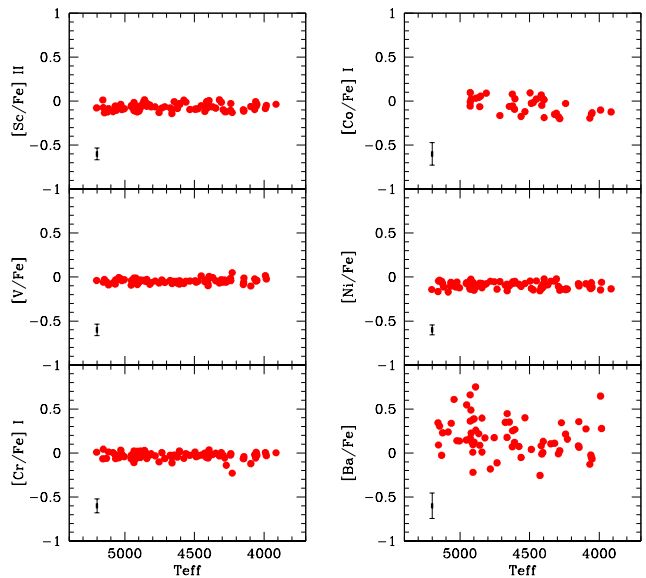}
\caption{Abundance ratios of elements of the Fe-peak (Sc~{\sc ii}, V, Cr, Co,
Ni) and the $s-$process element Ba from GIRAFFE spectra as a function of the
effective temperature. Error bars indicate internal star to star errors.}
\label{f:fegm36}
\end{figure}

The pattern of the $\alpha-$ and Fe-group element abundances is summarized in
Fig.~\ref{f:alpm36} and Fig.~\ref{f:fegm36} where the abundances are plotted as
a function of the effective temperature for individual stars in NGC~362.

In the last panel (bottom right) of Fig.~\ref{f:fegm36} we also show the
abundance ratios of Ba, which is a neutron-capture element whose abundance in
the solar system is mostly due to the $s-$process.  None of
these elements present a trend as a function of temperature, and as shown in
Table~\ref{t:meanabu36} their abundances are usually very homogeneous in
NGC~362, apart from the light elements involved in proton-capture reactions
discussed in the previous Section. The case of Ba with its large spread will be
discussed in detail in Section 5.4 below.

\begin{figure}
\centering 
\includegraphics[bb=19 145 580 519, clip, scale=0.42]{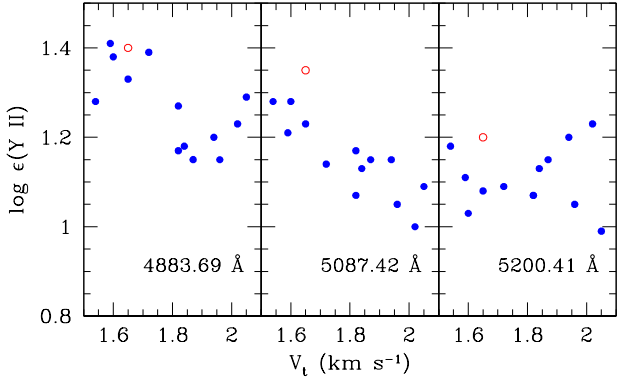}
\caption{$\log \epsilon$(Y~{\sc ii}) as derived from the three available clean
lines in our spectral range as a function of micro-turbulent  velocity. Star
11413 on the secondary, red RGB is indicated with a red open circle.}
\label{y:vm}
\end{figure}

\begin{figure}
\centering 
\includegraphics[scale=0.42]{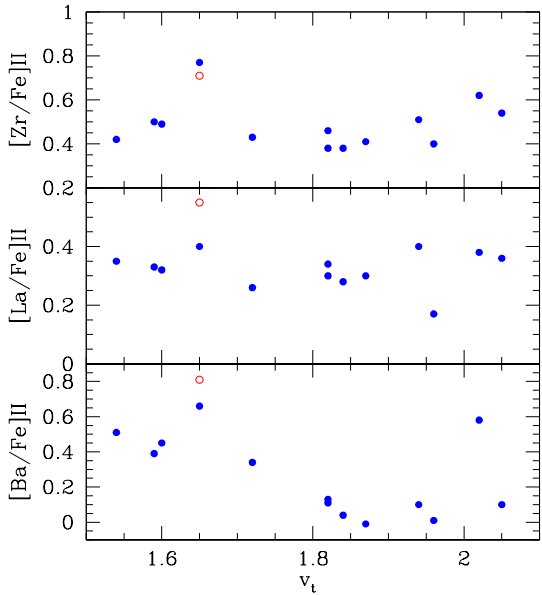}
\caption{Abundances of Zr, Ba and La as a function of micro-turbulent velocity
for stars with UVES spectra. The empty circle indicates star 11413 (see text).}
\label{ba:vm}
\end{figure}

Analysis of the UVES spectra allows the measurements of the abundances of
several more n-capture elements, including Cu, Y, La, Ce, Nd, Eu, and Dy, 
beside Ba which is measured also from GIRAFFE spectra. Since stars observed with
UVES are typically very cool and luminous, some of the relevant lines are strong
and then derived abundances are highly sensitive to microturbulence velocity.
For some lines we obtained strong correlation between abundances and the adopted
micro-turbulent velocities. As an example, Fig.~\ref{y:vm} shows the
correlations obtained for  three clean Y~{\sc ii} lines. The lines at 4883.7 and
5087.4~\AA\  show a marked  trend, therefore we only used the line at 
5200.4~\AA. With similar considerations, we selected only Nd lines with $EW$
smaller than 70~m\AA, yielding typically six lines per stars. In the case of Ba
the three available lines, 5853.7, 6141.7, 6469.9~\AA\ are all very strong, and
thus all sensitive to microturbulence. Fig.~\ref{ba:vm} shows  [Zr/Fe], [Ba/Fe],
and [La/Fe] as a function of microturbulent velocity. The trend of Ba with
microturbulence is quite obvious. However, part of the trend is due to star
11413, which is richer in all n-capture elements than the other giants in the
UVES sample. When this star is excluded, the trend between Ba abundances and
microturbulence values is scarcely significant, at less than two-sigma.

We note that in order to further flatten the trend of Ba (and of the  strong Nd
and Y lines) with microturbulence, the star-to-star variation of such  parameter
should be decreased, i.e. the low microturbulence values increased and  the high
ones decreased. However, this would create a trend of the Fe abundances with
microturbulence of the opposite sign. Given that the measured  Fe lines
typically form deeper in the stellar atmospheres than the strong lines  of heavy
elements, this opposite behaviour with respect to microturbulence hints  at an
intrinsic inadequacy of the model atmosphere for cool and luminous stars.  Note
that this effect is likely not as strong in those stars observed using GIRAFFE
(see Fig.~\ref{f:fegm36}), as they are typically warmer and fainter.  Therefore,
the derived errors for Ba indicated in Table A.1 for the UVES stars  should be
taken with caution, as the true errors are likely larger, due to the  very high
sensitivity to the adopted microturbulent velocity. Finally, we note  that an
even stronger trend of Ba abundances with microturbulent velocity is  present in
the analysis by Shetrone and Keane (2000), albeit in that case Ba  abundances
increase as $V_t$ increases.

\begin{figure}
\centering 
\includegraphics[scale=0.42]{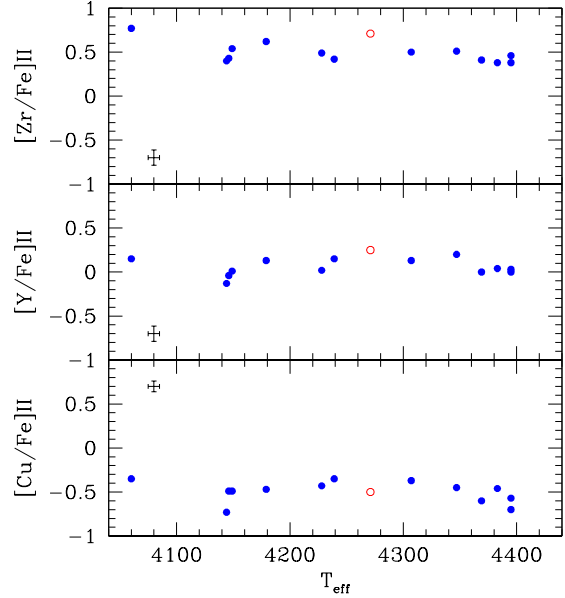}
\caption{Abundance ratios of elements of the n-capture elements Zr~{\sc ii},
Y~{\sc ii}, and  Cu~{\sc i}, from UVES spectra as a function of the effective
temperature. Error bars indicate internal star-to-star errors. In each panel,
the Ba-rich star 11413 is highlighted.}
\label{n1:teff}
\end{figure}

\begin{figure}
\centering 
\includegraphics[scale=0.42]{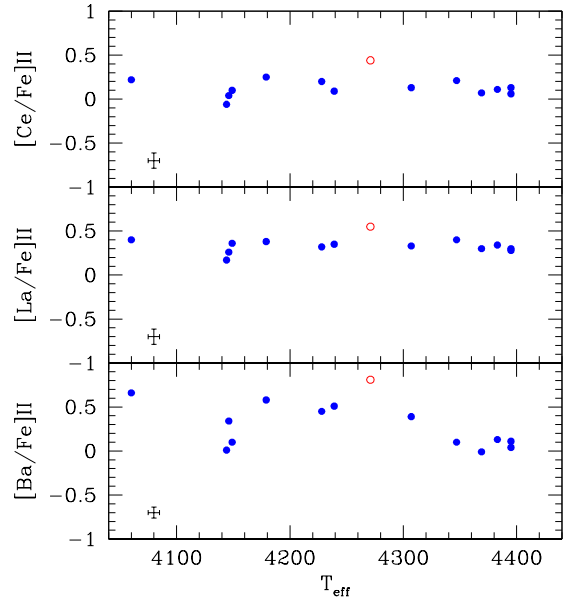}
\caption{As in Fig.~\ref{n1:teff} for the n-capture elements Ba~{\sc ii}, 
La~{\sc ii}, and Ce~{\sc ii}.}
\label{n2:teff}
\end{figure}

\begin{figure}
\centering 
\includegraphics[scale=0.42]{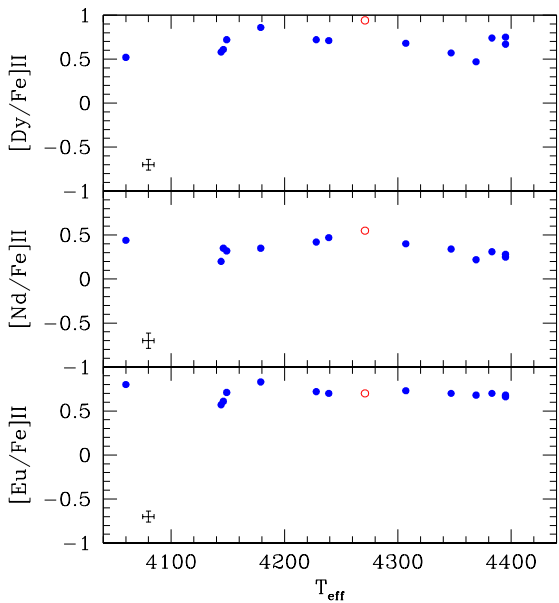}
\caption{As in Fig.~\ref{n1:teff} for  the n-capture elements Eu~{\sc ii}, 
Nd~{\sc ii}, and Dy~{\sc ii}.}
\label{n3:teff}
\end{figure}

Fig.~\ref{n1:teff}, Fig.~\ref{n2:teff}, and Fig.~\ref{n3:teff} show the derived 
abundances for the heavy elements from UVES spectra as a function of $T_{\rm
eff}$.   There is a weak trend of Ba abundances with $T_{\rm eff}$ due to the
above discussed  correlation between derived Ba abundance and microturbulent
velocity, given that the latter has a trend with effective temperature. Apart
from the case of Ba, the measured n-capture elements abundances are quite
uniform, showing remarkably small scatter, especially when star 11413 is
excluded from the average. This giant has the highest abundance of Y, Ba,
La, Nd, Ce, and Dy among the sample of 14 stars with UVES spectra, whereas its
Eu abundance does not stand out among the other stars in this sample.

\subsection{$s-$ and $r-$process contributions}

The abundances of neutron capture elements derived from UVES spectra can be used
to estimate the relative weight of the $s-$ and $r-$process contributions to the
chemical pattern in NGC~362, as done in Carretta et al. (2011) for NGC~1851 and 
in D'Orazi et al. (2011) for $\omega$ Cen. We refer to those papers for the
details on the adopted procedure.
Results for Ba, La, Ce, and Eu are well consistent with each other: 
the pattern of abundances is reproduced if we assume that most of these elements
are made by the $r-$process (1/3 of the solar $r-$process abundances) with a small 
contribution from the $s-$process (about 1/25 of the solar $s-$process abundances).

On the other hand, results for the first-peak elements Y and Zr are discrepant:
we found less Y and too much Zr with respect to the above recipe. The
uncertainties in the determination of Y and Zr do not allow to determine with
precision the pattern of the first peak species. Since we have well defined
abundances only for the second peak, we are unable to calculate the ratio
between heavy and light $s-$process elements in NGC~362. Hence, nothing  more
can be  said about the origin of these elements; in turn, the timescales of the
enrichment cannot be properly determined.

\subsection{Ba and the red sequence on the RGB}

\begin{figure}
\centering 
\includegraphics[scale=0.40]{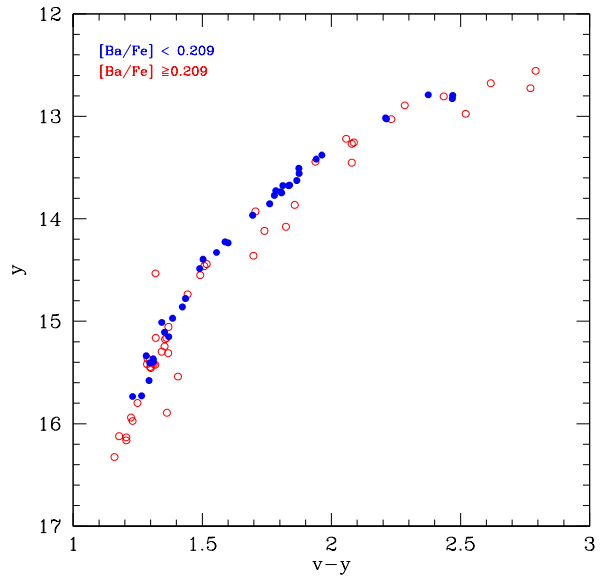}
\caption{Str\"omgren $y,v-y$ CMD for stars of our sample with accurate
photometry. The stars are divided according to their Ba abundances: blue filled
circles are stars with [Ba/Fe] lower than the average value [Ba/Fe]=+0.209 dex,
whereas red open circles indicate stars with Ba content larger than the mean
value.}
\label{f:bavy36}
\end{figure}

The Ba abundances for the about 70 stars observed with GIRAFFE  are based on a 
single line at 6141~\AA, and the associated internal error is 
rather large ($\sim 0.15$~dex). However, the $r.m.s.$ scatter of the mean
(0.21 dex) is larger than the internal errors (see Fig.~\ref{f:fegm36}, 
bottom right panel, and Fig.~\ref{n2:teff}). Additional evidence that the 
spread may be real, albeit small, comes from the Str\"omgren photometry. We 
used the photometry collected by Grundahl and coworkers, and presented by 
Calamida et al. (2007)\footnote{The catalogue was downloaded from the web 
page http://www.oa-roma.inaf.it/spress/gclusters.html} to whom we refer for 
details. There are 69 stars having both photometric data and abundances of Ba. 
We show the Str\"omgren $y,v-y$ CMD for these stars in Fig.~\ref{f:bavy36}, 
where we indicated with different symbols stars with Ba abundances lower or 
higher than the average value of [Ba/Fe]=+0.209 dex. Stars with lower than
average Ba abundances define a very narrow sequence at the blue ridge of the 
RGB, while a fraction of the Ba-rich stars populate a sequence slightly to
the red of the main RGB (apart from an outlier with a  very blue $v-y$\
colour likely due to errors in the photometry). We will come back to this
red sequence in Section~6.

We note that the above discussed uncertainty in the Ba abundance due to
microturbulence does not affect this result. In fact, at a given effective
temperature T$_{eff}$, the relative Ba abundances are robust with respect to microturbulence,
therefore at a given magnitude the separation between Ba-rich and Ba-poor stars
is reliable.

\begin{figure}
\centering 
\includegraphics[scale=0.40]{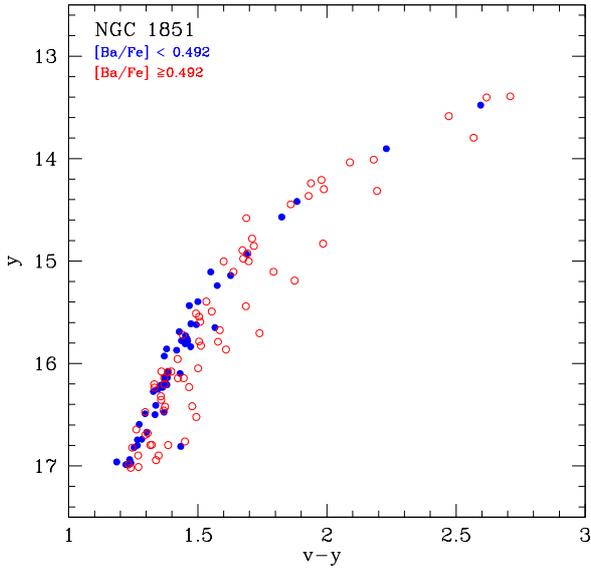}
\caption{Str\"omgren $y,v-y$ CMD for stars of NGC~1851 from Carretta et al.
(2011). The stars are divided according to their Ba abundances: blue filled
circles and red open circles are stars with [Ba/Fe] lower and higher,
respectively than the average value [Ba/Fe]=+0.492 dex.}
\label{f:bavy18}
\end{figure}

This behaviour reproduces the splitting of the RGB of NGC~1851 observed in the 
$v-y$ colours, as shown in Fig.~\ref{f:bavy18} using data from Carretta et al. 
(2011). In both cases, the reddest sequence is populated entirely by Ba-rich 
stars.

What is the physical explanation for this phenomenon? In Carretta et al. (2011)
we suggested that the redder $v-y$\ colour of this secondary sequence
is most likely due to large enhancements in N. Very recently,
this suggestion was supported by the analysis of N abundances based on
a number of CN features measured on GIRAFFE spectra (Carretta et al. 2013,
in preparation), as well as on a limited sample of giants by Villanova et al.
(2010) in NGC~1851: the stars populating the reddest sequence all have
strong CN bands and higher N abundances.

The low resolution data for bright giants by Smith \& Langland-Shula (2009) may 
be used to check if the same holds for NGC~362. These authors provided values of 
the index S(3839) which measure the strength of the CN band strength. These
values were converted into the quantity $\delta$S(3839) as defined by Norris
(1981) using for the baseline of equation S(3839)$=0.44+0.0878\times M_V$ based
on the data in Smith \& Langland-Shula (2009). Note that the stars in common
between the two samples span a very limited range in magnitude and colour along
the RGB of NGC~362 ($-2.23<M_V<-1.92$). Hence, any variation of the
$\delta$S(3839) index with luminosity and/or temperature has no impact in our
discussion. 

\begin{figure}
\centering 
\includegraphics[scale=0.40]{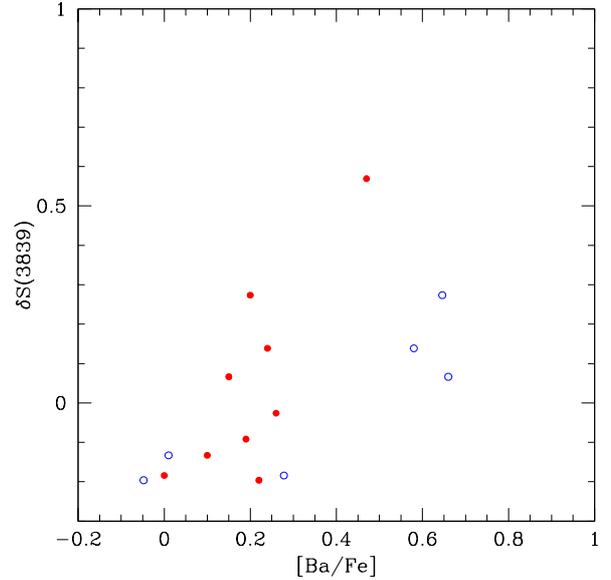}
\caption{The index S(3839) of the CN bandstrength from Smith and Langland-Shula
(2009) as a function of the [Ba/Fe] ratio from Shetrone and Keane (2000, red
filled point) and from the present work (blue open circles) for stars of
NGC~362.}
\label{f:s3839ba}
\end{figure}

We found 
that, beside to the already known correlation with Na and anti-correlation with 
O abundances, this index is also well correlated with the [Ba/Fe] ratio (we
notice that the same effect was found in M~22 by Marino et al. 2011). This is
shown in Fig.~\ref{f:s3839ba}, where we adopted Ba abundances from both the few
stars of the present study and the more numerous giants in common with Shetrone
and Keane (2000). Among the six stars in our sample, star 2333, with the
highest value of $\delta$S(3839), lies on the red RGB sequence.

\begin{figure}
\centering 
\includegraphics[scale=0.40]{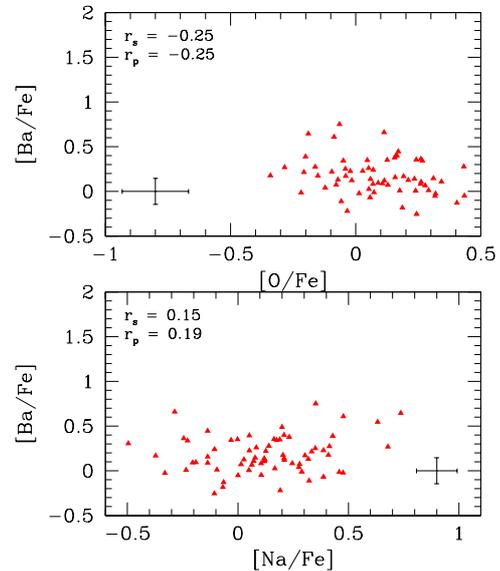}
\caption{Upper panel: [Ba/Fe] as a function of [O/Fe] in giants of NGC~362 from
this study. Lower panel: the same, as a function of [Na/Fe]. The Pearson linear
correlation coefficient and the Spearman  rank correlation coefficient, together
with star to star error bars, are  indicated in each panel.}
\label{f:banamg36a}
\end{figure}

\begin{figure}
\centering 
\includegraphics[scale=0.40]{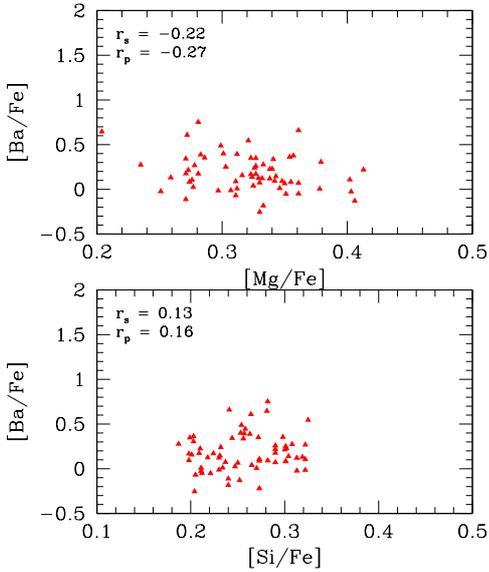}
\caption{As in the previous figure, but for [Mg/Fe] (upper panel) and [Si/Fe]
(lower panel).}
\label{f:banamg36b}
\end{figure}

Since Ba is correlated with Na and anti-correlated with O in this subset of stars, 
we checked that the same also holds in our much larger sample observed with
GIRAFFE.
In Figs.~\ref{f:banamg36a} and \ref{f:banamg36b} we show the run of Ba
abundances as a function of the content of O, Na, Mg, and Si. The best 
(anti-)correlations are those between Ba and the elements that are depleted in 
proton-capture reactions, like O and Mg. The linear correlation
coefficients are $r=-0.25$ and $-0.27$, respectively (with 61 and 62
degrees of freedom, respectively). These
anti-correlations, although not striking, are significant to a level of 
confidence of about 95\%, admittedly not very high, while the 
complementary correlations with Na and Si are not statistically significant.

At the moment, the reason why a typical $s-$process element like Ba must be
related to the abundances of elements forged in the nuclear processing of likely
much more massive stars of the first stellar generation in the cluster still 
remains an unsolved issue, owing to the much different lifetimes of typical
polluters for these elements. More efforts should be focused in the future
in ascertaining the ground of these possible correlations in large samples of
stars.

\begin{figure}
\centering 
\includegraphics[bb=19 140 589 474, clip, scale=0.45]{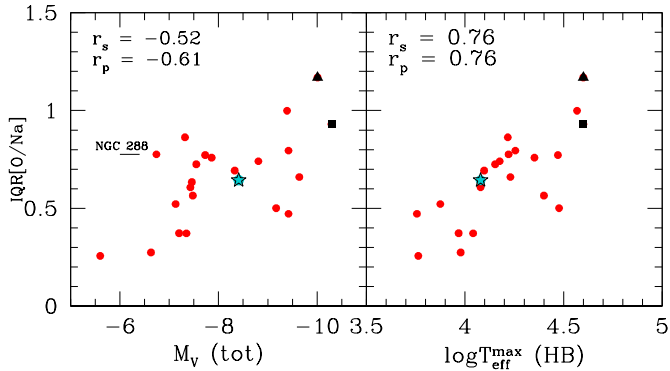}
\caption{Interquartile range (IQR) of the [O/Na] ratio in NGC~362 as a function
of the total absolute magnitude of the cluster (left panel), and of the maximum
temperature along the HB (right panel). NGC~362 is represented by a filled star
symbol. The filled square is for $\omega$ Cen from Johnson and Pilachowski 
(2010) and the filled triangle indicates M~54 (Carretta et al. 2010c). The other
clusters are from Carretta et al. (2009a, 2011). In the left panel, the location
of NGC~288 is indicated. In each panel the Spearman rank correlation coefficient
($r_s$) and the Pearson's correlation coefficient are reported.} 
\label{f:iqrmvteff36}
\end{figure}

\section{Discussion and conclusions}

We studied the chemical composition of 92 red giant branch stars in the
globular cluster NGC~362. 

We found that NGC~362 behaves like a perfectly 'normal' cluster for what
concerns the elements involved in high temperature H-burning. It nicely falls 
in the middle of the relation between the extension of the Na-O anti-correlation
and the total absolute magnitude (a proxy for the present-day cluster
mass) defined by the  other GCs (Carretta et al. 2010a; see left panel of
Fig.~\ref{f:iqrmvteff36}).  The same holds for the correlation of the extension
of Na-O anti-correlation with the extent of the HB distribution (right panel of
Fig.~\ref{f:iqrmvteff36}).

In NGC~362 there is a hint that stars are clustered into two discrete
groups along the Na-O anti-correlation: this comes both from the more
conspicuous sample of RGB stars with GIRAFFE spectra and the more limited
fraction of stars with high-resolution FLAMES/UVES spectra.

First-generation stars (Na-poor and O-rich) are apparently less concentrated
than second generation stars, in agreement with the result found in most
GCs both from spectroscopy (e.g. Carretta et al. 2010d) and photometry (e.g.
Lardo et al. 2011, Milone et al. 2012 and references therein); however, this
finding is not supported by a high level of statistical confidence  in NGC~362.

In this context, it is interesting to compare NGC~362 to its twin
second-parameter cluster NGC~288 (Shetrone and Keane 2000, and, in particular, 
Carretta et al. 2009a,b). The latter GC is indicated in Fig.~\ref{f:iqrmvteff36}
(left panel) to show its position slightly off the main relation. Carretta et
al. (2010a) already discussed evidence (like tidal tails) pointing toward a
large loss of stars for NGC~288 in the past. Another evidence in this sense
is given by the study by Paust et al. (2010), who determined the global present
day mass functions for 17 GCs, including NGC~288 and NGC~362. Their comparison
shows that NGC~288 has a flatter mass function (the slope of the power law fit
is $\alpha=-0.83$, compared to $\alpha=-1.69$ of NGC~362), indicating a
lower fraction of low mass stars. Since NGC~288 is a more loose cluster than its
twin NGC~362, this seems to imply that is also more affected by external
processes of evaporation through disk shocking and/or tidal stripping (see Paust
et al. 2010), well explaining  its position in the plane IQR[O/Na] vs $M_V$.

Also relevant in the context of multiple populations in GCs is the recent
addition of NGC~362 to the increasing number of clusters where a broad and/or
even split subgiant branch (SGB) was detected. The studies by Milone and
collaborators include: NGC~1851 (Milone et al. 2008), 47 Tuc (Anderson et al.
2009, Milone et al. 2012), NGC~6388 and (maybe) NGC~6441 (Bellini et al. 2013), 
NGC~362, NGC~5286, NGC~6656, NGC~6715, and NGC~7089 (Piotto et al. 2012).

Our FLAMES survey, devoted to study the abundances of proton-capture elements
(mainly Na and O) in large samples of RGB stars in a large number of GCs, is
blind to the feature of double SGBs, which is governed by age and/or CNO content
(see e.g. Cassisi et al. 2008). However, as already noted in previous section,
a further peculiarity of NGC~362 is its {\it secondary giant branch} in CMD
using the $v-y$ color (see Fig.~\ref{f:bavy36}).

The reality of the sequence and the membership of this population to the
cluster  is well assessed in the present work. Our spectroscopic analysis shows
that all  the stars on this {\it secondary giant branch} have measured RV and
metallicity consistent with the cluster. This sequence is neither grossly
contaminated by Galactic field interlopers, nor by stars of the Small Magellanic
Cloud, found at fainter magnitudes and well separated from the stars in NGC~362
in the $y,v-y$ plane.
Thus, beside the split SGB, NGC~362 seems to be part  of the group of GCs  where
a {\it secondary giant branch} is found in the $y,v-y$ plane.  The most relevant
cases  include NGC~1851, NGC~6656 (M~22), and NGC~7089 (M~2: Lardo et al. 2012).
With the  addition of NGC~362 the phenomenon becomes common enough to justify
the dubbing of  ``secondary RGB" for this feature, since it seems to enclose
only a minor fraction  of the total red giants, although the percentage does
vary from cluster to cluster.

In the present study we showed that the ``secondary RGB" of NGC~362 is only
populated  by stars with high Ba abundance, compared to the average of the
sample.\footnote{It also seems to show a tendency for lower O and higher Na
abundances, although we measured both elements only in six stars. This will be
discussed more in a companion paper.}  The conspicuous  blue RGB includes
instead both Ba-rich and Ba-normal stars. Is this separation only  restricted to
the Ba abundances? Star 11413 is the only star on this sequence observed  with
UVES. 
Unfortunately, this star was classified as binary by
F93. This would have no impact on the abundance analysis,
unless the binary system underwent mass transfer from an
AGB companion, a rare event. While this occurrence would
explain the kinematics and part of the chemical pattern of
this star, surely it is not the explanation of the secondary
sequence: were the stars on this sequence this kind of
binaries, we would expect a range in variations (also
in colour), not a sequence. Anyway, to be conservative, we
cannot infer from this star alone that the enhancement observed
for Ba is extended to all the $s-$process elements.

This pattern with the red sequence only populated by stars with high Ba
abundances is reminiscent of the chemical characterization of the red RGB in
NGC~1851 that are found to be preferentially enriched in $s-$process elements (and in
particular, Ba) in a number of recent studies (Yong and Grundahl 2008, Villanova
et al. 2010, Carretta et al. 2011)\footnote{The small sample by Yong and
Grundahl is not assigned to one of the RGBs; however, in Fig.1 of Lee et al.
(2009) three s-rich stars are seen to lie on the reddest branch.}.
The same pattern, even more marked, is present in M~22,  where however stars on
the bluer RGB are all $s-$poor and stars on the red RGB are all $s-$rich, as
demonstrated by Marino et al. (2009, 2011). In that cluster, the separation is
more neat.

\begin{figure}
\includegraphics[bb=20 149 297 705, clip, scale=0.8]{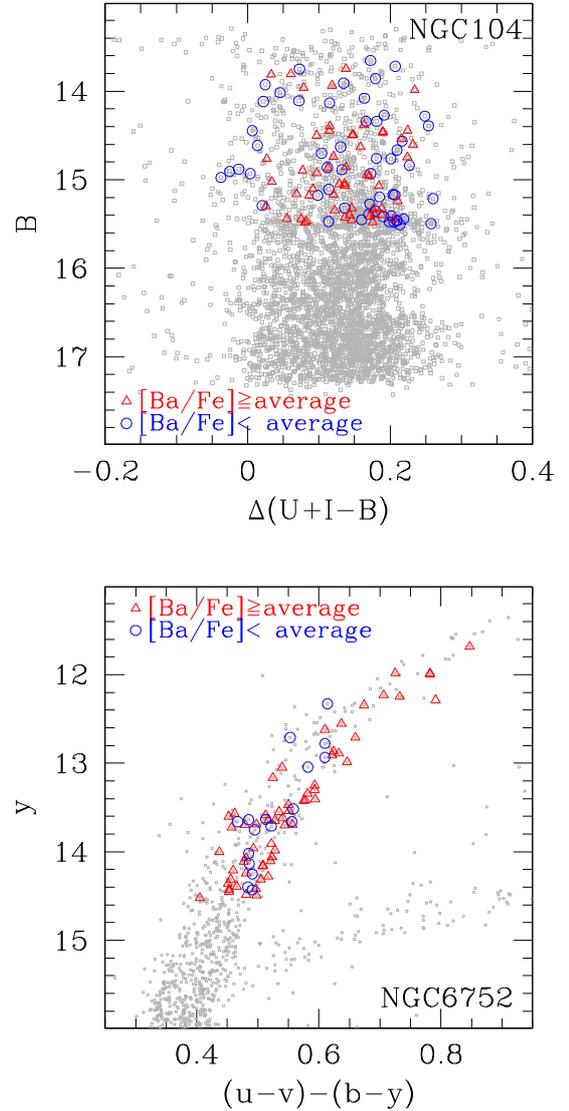}
\caption{Top: the plot shows 47~Tuc, using the same filter combination adopted
in Milone et al. (2012) to show a split RGB (see their Figs.~21 and 26); the
photometry is not the same. Ba-rich and Ba-poor stars are not segregated.
Bottom: for NGC~6752 we show the same effect on a plot similar to the one in
Milone et al. (2013; their Fig.~15).}
\label{f:ba4767} 
\end{figure}

A more detailed discussion of the properties of this ``secondary RGB" in NGC~362
and a comparison with other peculiar 
GCs  will be presented in a separate  paper (Carretta et al. 2013b). Here we
only note that the splitting in the $y,v-y$\ diagram seems to be different from
the split RGB found in clusters such as, e.g., NGC~104 (47~Tuc) or NGC~6752,
when using colours  including ultraviolet magnitudes, which are closely related
to the Na-O anti-correlation  (see the extensive photometric studies by Milone
et al. 2012, 2013 and the theoretical interpretation by Sbordone et al. 2011).
Not all splits in the different filter combinations have the same meaning (see
the discussion in Carretta et al. 2013b and Milone et al. 2012). No
``secondary RGB"  is easily discernable in the $y,v-y$\ diagrams for the two GCs
just mentioned (at least, using the photometry by Calamida et al. 2007) showing
that this feature is present in some, but not all GCs.  These GCs also behave
differently for what concerns Ba abundances. To show this, we used the  Johnson
UBVI photometry by Momany (unpublished photometry, reduced as in Momany et al.
2003) for 47~Tuc and the Str\"omgren photometry by Grundahl for NGC~6752
(Calamida et al. 2007), to reproduce the clear splits shown by Milone et al.
(2012,2013). We took the Ba abundances from D'Orazi  et al. (2010),
cross-identified the stars,  and plotted the ones in common in
Fig.~\ref{f:ba4767}. When we divide the sample in Ba-rich or Ba-poor stars with
respect to the average [Ba/Fe] value in each cluster (as done in NGC~1851 by
Carretta et al. 2011 and in the present work for NGC~362), no segregation is
evident on the split RGBs in these two clusters, at odds with what is seen for
C,N and Na,O (Milone et al. 2012, 2013).

In the last years we are learning new things on the nature of GCs, which are more complex -and more interesting- than we thought for long time. True progress and understanding requires a multi-lateral approach, making the best of the information that photometry, spectroscopy, and theoretical modelling can give us.
                    
\begin{acknowledgements}
VD is an ARC Super Science Fellow.
This publication makes use of data products from the Two Micron All Sky Survey,
which is a joint project of the University of Massachusetts and the Infrared
Processing and Analysis Center/California Institute of Technology, funded by the
National Aeronautics and Space Administration and the National Science
Foundation.  This research has been funded by PRIN INAF 2011
"Multiple populations in globular clusters: their role in the Galaxy assembly"
(PI E. Carretta), and PRIN MIUR 2010-2011, project ``The Chemical and Dynamical
Evolution of the Milky Way and Local Group Galaxies'' (PI F. Matteucci) . This
research has made use of the SIMBAD database, operated at CDS, Strasbourg,
France and of NASA's Astrophysical Data System.
\end{acknowledgements}

\begin{appendix}

\section{Error estimates}

We refer the reader to the analogous Appendices in Carretta et al. (2009a,b) for
a detailed discussion of the procedure adopted for error estimates. 
In the following we only provide the main tables of sensitivities of abundance
ratios to the adopted errors in the atmospheric parameters and $EW$s and the
final estimates of internal and systematic errors for all species analysed from
UVES and GIRAFFE spectra of stars in NGC~362.

The sensitivities of derived  abundances on the adopted atmospheric parameters 
were obtained by repeating our abundance analysis by changing only one 
atmospheric parameter each time for $all$\ stars in NGC~362 and taking the 
average value of the slope change vs. abundance. This exercise was done separately 
for both UVES and GIRAFFE spectra.

We notice that when estimating the contribution to internal errors due to EWs
and $v_t$, the values usually adopted (determined from the scatter of abundances 
from individual lines) are overestimated, because regularities in
the data are not taken into account. These regularities are due to
uncertainties in the $gf-$values, unrecognised blends with adjacent lines,
not appropriate positioning of the continuum, etc. They show up in uniform deviations
of individual lines from average abundances for each star. By averaging
over all stars the residuals of abundances derived from individual lines with 
respect to the average value for each star, we estimated that some 36\% of 
the total variance in the Fe abundances from individual lines is due to 
systematic offsets between different lines, which repeat from star-to-star. 
For about 30\% of the line, these offsets have trends with temperature
significant at about 2~$\sigma$ level. However, we found that the additional 
fraction of variance that can be explained by these trends is very small, and 
we can neglect it. We conclude that when considering star-to-star variations 
(internal errors, according to our denomination), the errors in EWs and $v_t$ 
should be multiplied by 0.8. 

The amount of the variations in the atmospheric parameters is shown in the
first line of the headers in Table~\ref{t:sensitivityu36},
and Table~\ref{t:sensitivitym36}, whereas the resulting 
response in abundance changes of all elements (the sensitivities) are shown 
in columns from 3 to 6 of these tables.

\begin{table*}
\centering
\caption[]{Sensitivities of abundance ratios to variations in the atmospheric
parameters and to errors in the equivalent widths, and errors in abundances for
stars in NGC 362 observed with UVES}
\begin{tabular}{lrrrrrrrr}
\hline
Element     & Average  & T$_{\rm eff}$ & $\log g$ & [A/H]   & $v_t$    & EWs     & Total   & Total      \\
                   & n. lines &      (K)      &  (dex)        & (dex)   &kms$^{-1}$& (dex)   &Internal & Systematic \\
\hline        
Variation&             &  50           &   0.20   &  0.10   &  0.10    &         &         &            \\
Internal &             &   5           &   0.04   &  0.05   &  0.04    & 0.087   &         &            \\
Systematic&            &  59           &   0.06   &  0.05   &  0.01    &         &         &            \\
\hline
$[$Fe/H$]${\sc  i}& 81 &    +0.041     &   +0.016 &  +0.003 & $-$0.033 & +0.010  &0.017    &0.051      \\
$[$Fe/H$]${\sc ii}&  9 &  $-$0.051     &   +0.104 &  +0.034 & $-$0.012 & +0.029  &0.041    &0.072      \\
$[$O/Fe$]${\sc  i}&  2 &  $-$0.029     &   +0.067 &  +0.032 &   +0.030 & +0.062  &0.066    &0.068      \\
$[$Na/Fe$]${\sc i}&  3 &    +0.008     & $-$0.048 &$-$0.032 &   +0.019 & +0.050  &0.054    &0.053      \\
$[$Mg/Fe$]${\sc i}&  3 &  $-$0.010     & $-$0.015 &$-$0.006 &   +0.018 & +0.050  &0.051    &0.017      \\
$[$Al/Fe$]${\sc i}&  2 &    +0.003     & $-$0.024 &$-$0.010 &   +0.028 & +0.062  &0.063    &0.052      \\
$[$Si/Fe$]${\sc i}&  9 &  $-$0.053     &   +0.023 &  +0.010 &   +0.025 & +0.029  &0.032    &0.064      \\
$[$Ca/Fe$]${\sc i}& 16 &    +0.021     & $-$0.030 &$-$0.018 & $-$0.016 & +0.022  &0.025    &0.027      \\
$[$Sc/Fe$]${\sc ii}& 7 &    +0.042     & $-$0.023 &$-$0.003 & $-$0.016 & +0.033  &0.034    &0.052      \\
$[$Ti/Fe$]${\sc i}&  9 &    +0.054     & $-$0.019 &$-$0.018 & $-$0.001 & +0.029  &0.031    &0.065      \\
$[$Ti/Fe$]${\sc ii}& 1 &    +0.030     & $-$0.019 &$-$0.005 &   +0.004 & +0.087  &0.087    &0.038      \\
$[$V/Fe$]${\sc i} &  9 &    +0.068     & $-$0.015 &$-$0.013 &   +0.001 & +0.029  &0.031    &0.081      \\
$[$Cr/Fe$]${\sc i}&  2 &    +0.037     & $-$0.020 &$-$0.028 & $-$0.010 & +0.062  &0.064    &0.048      \\
$[$Cr/Fe$]${\sc ii}&16 &    +0.014     & $-$0.025 &$-$0.020 &   +0.008 & +0.022  &0.025    &0.026      \\
$[$Mn/Fe$]${\sc i}&  3 &    +0.021     & $-$0.013 &$-$0.009 & $-$0.018 & +0.050  &0.051    &0.027      \\
$[$Co/Fe$]${\sc i}&  5 &  $-$0.008     &   +0.002 &  +0.003 &   +0.024 & +0.039  &0.040    &0.019      \\
$[$Ni/Fe$]${\sc i}& 29 &  $-$0.014     &   +0.016 &  +0.008 &   +0.014 & +0.016  &0.018    &0.019      \\
$[$Cu/Fe$]${\sc i}&  1 &    +0.010     & $+$0.019 &$-$0.003 & $-$0.017 & +0.100  &0.103    &0.091      \\
$[$Y/Fe$]${\sc ii}&  2 &    +0.040     & $-$0.037 &$-$0.006 &   +0.008 & +0.100  &0.087    &0.050      \\
$[$Zr/Fe$]${\sc ii}& 2 &    +0.041     & $-$0.017 &$-$0.005 &   +0.002 & +0.040  &0.087    &0.067      \\
$[$Ba/Fe$]${\sc ii}& 3 &    +0.071     & $-$0.030 &$-$0.002 & $-$0.088 & +0.010  &0.062    &0.097      \\
$[$La/Fe$]${\sc ii}& 1 &    +0.051     & $-$0.071 &  +0.000 & $-$0.001 & +0.100  &0.087    &0.073      \\
$[$Ce/Fe$]${\sc ii}& 2 &    +0.041     & $-$0.017 &$-$0.006 &   +0.002 & +0.050  &0.087    &0.071      \\
$[$Nd/Fe$]${\sc ii}& 7 &    +0.065     & $-$0.021 &$-$0.001 &   +0.008 & +0.042  &0.087    &0.065      \\
$[$Eu/Fe$]${\sc ii}& 2 &    +0.051     & $-$0.037 &$-$0.001 &   +0.001 & +0.062  &0.062    &0.055      \\
$[$Dy/Fe$]${\sc ii}& 1 &    +0.031     & $-$0.024 & $-$0.004 &  +0.002 & +0.050  &0.062    &0.097      \\
\hline
\end{tabular}
\label{t:sensitivityu36}
\end{table*}

\begin{table*}
\centering
\caption[]{Sensitivities of abundance ratios to variations in the atmospheric
parameters and to errors in the equivalent widths, and errors in abundances for
stars in NGC 362 observed with GIRAFFE}
\begin{tabular}{lrrrrrrrr}
\hline
Element     & Average  & T$_{\rm eff}$ & $\log g$ & [A/H]   & $v_t$    & EWs     & Total   & Total      \\
            & n. lines &      (K)      &  (dex)   & (dex)   &kms$^{-1}$& (dex)   &Internal & Systematic \\
\hline        
Variation&             &  50           &   0.20   &  0.10   &  0.10    &         &         &            \\
Internal &             &   5           &   0.04   &  0.04   &  0.12    & 0.126   &         &            \\
Systematic&            &  59           &   0.06   &  0.06   &  0.01    &         &         &            \\
\hline
$[$Fe/H$]${\sc  i}& 29 &   +0.052      &   +0.000 &$-$0.005 & $-$0.028 & +0.023  &0.041    &0.062      \\
$[$Fe/H$]${\sc ii}&  2 & $-$0.033      &   +0.091 &  +0.023 & $-$0.011 & +0.089  &0.093    &0.048      \\
$[$O/Fe$]${\sc  i}&  1 & $-$0.041      &   +0.081 &  +0.034 &	+0.031 & +0.126  &0.133    &0.059      \\
$[$Na/Fe$]${\sc i}&  2 & $-$0.011      & $-$0.031 &$-$0.012 &	+0.018 & +0.089  &0.092    &0.031      \\
$[$Mg/Fe$]${\sc i}&  2 & $-$0.018      & $-$0.007 &$-$0.002 &	+0.016 & +0.089  &0.091    &0.022      \\
$[$Si/Fe$]${\sc i}&  6 & $-$0.046      &   +0.025 &  +0.010 &	+0.027 & +0.051  &0.061    &0.055      \\
$[$Ca/Fe$]${\sc i}&  5 &   +0.007      & $-$0.024 &$-$0.006 & $-$0.010 & +0.056  &0.058    &0.011      \\
$[$Sc/Fe$]${\sc ii}& 4 & $-$0.053      &   +0.080 &  +0.030 &	+0.011 & +0.063  &0.068    &0.067      \\
$[$Ti/Fe$]${\sc i}&  3 &   +0.023      & $-$0.008 &$-$0.008 &	+0.009 & +0.073  &0.074    &0.027      \\
$[$V/Fe$]${\sc i} &  4 &   +0.037      & $-$0.006 &$-$0.008 &	+0.011 & +0.063  &0.065    &0.044      \\
$[$Cr/Fe$]${\sc i}&  3 & $-$0.029      & $-$0.045 &$-$0.043 & $-$0.021 & +0.073  &0.080    &0.037      \\
$[$Co/Fe$]${\sc i}&  1 &   +0.005      &   +0.011 &  +0.007 &	+0.022 & +0.126  &0.129    &0.018      \\
$[$Ni/Fe$]${\sc i}&  6 & $-$0.010      &   +0.015 &  +0.006 &	+0.020 & +0.051  &0.057    &0.014      \\
$[$Ba/Fe$]${\sc ii}& 1 & $-$0.038      &   +0.057 &  +0.037 & $-$0.057 & +0.126  &0.145    &0.055      \\
\hline
\end{tabular}
\label{t:sensitivitym36}
\end{table*}

\end{appendix}

\clearpage

\setcounter{table}{1}
\begin{table*}
\centering
\caption{List and relevant information for target stars in NGC~362
The complete
Table is available electronically only at CDS.}
\begin{tabular}{rllcccrl}
\hline
    ID    &RA           &Dec          &   $B$ &   $V$ &   $K$  &RV(Hel)&Notes     \\
\hline
     986  &  1 03 20.037& -70 49 55.53& 15.227& 14.139 & 11.443 & 228.83 & HR11,HR13  \\  
    995 & 1 03 23.159 & -70 49 54.98 & 15.446 & 14.548 & 12.013 & 222.77 & HR13       \\  
\hline
\end{tabular}
\label{t:coo36}
\end{table*}

\clearpage
\setcounter{table}{2}
\begin{table*}
\centering
\caption[]{Adopted atmospheric parameters and derived iron abundances. The
complete table is available electronically only at CDS.}
\begin{tabular}{rccccrcccrccc}
\hline
Star   &  $T_{\rm eff}$ & $\log$ $g$ & [A/H]  &$v_t$	     & nr & [Fe/H]{\sc i} & $rms$ & nr & [Fe/H{\sc ii} & $rms$ \\
       &     (K)	&  (dex)     & (dex)  &(km s$^{-1}$) &    & (dex)	  &	  &    & (dex)         &       \\
\hline
  986  & 4533 &1.59& -1.18& 2.00&   52& $-$1.176& 0.161 &  3&  $-$1.141& 0.164 \\  
  995  & 4661 &1.92& -1.26& 1.09&   28& $-$1.259& 0.111 &  3&  $-$1.154& 0.028 \\  
\hline
\end{tabular}
\label{t:atmpar36}
\end{table*}

\clearpage
\setcounter{table}{3}
\begin{table*}
\centering
\caption[]{Abundances of proton-capture elements in stars of NGC~362.
n is the number of lines used in the analysis. Upper limits (limO,Al=0)
and detections (=1) for O and Al are flagged.}  
\begin{tabular}{rrccrccrccrcccc}
\hline
       star  &n &  [O/Fe]&  rms  &  n& [Na/Fe]& rms  &	n& [Mg/Fe]& rms  &  n&[Al/Fe]& rms  &limO&    limAl  \\ %
\hline       
  986  &   1 &	+0.16 &      &	 4 &  +0.21 & 0.12 &   3 & +0.30 &  0.16 &     &       &       &  1  &    \\  
  995  &   1 &	+0.17 &      &	 2 &$-$0.14 & 0.04 &     &	 &	 &     &       &       &  1  &    \\  
\hline
\end{tabular}
\label{t:proton36}
\end{table*}

\clearpage
\setcounter{table}{4}
\begin{table*}
\centering
\caption[]{Abundances of $\alpha$-elements in stars of NGC~362. 
n is the number of lines used in the analysis.}
\begin{tabular}{rrccrccrccrcc}
\hline
   star      &  n&[Si/Fe]&  rms &    n &  [Ca/Fe]& rms &   n &[Ti/Fe]~{\sc i} &  rms &n &[Ti/Fe]~{\sc ii} & rms \\
\hline   
  986  &   10 & +0.25 & 0.15 &   5 & +0.33 & 0.16 &   5 & +0.20 &  0.11 &      &       &  \\  
  995  &    1 & +0.26 &      &   6 & +0.32 & 0.23 &	&	&	&      &       &  \\  
\hline
\end{tabular}
\label{t:alpha36}
\end{table*}

\clearpage
\setcounter{table}{5}
\begin{table*}
\centering
\caption[]{Abundances of Fe-peak elements in stars of NGC~362. 
n is the number of lines used in the analysis.}
\scriptsize
\setlength{\tabcolsep}{1.3mm}
\begin{tabular}{rrccrccrccrccrccrccrcc}
\hline
      star    & n &[Sc/Fe]~{\sc ii}&rms&n& [V/Fe]  & rms  &  n &[Cr/Fe]~{\sc i}&rms&n& [Mn/Fe] & rms  &   n &[Co/Fe] & rms   &  n  &[Ni/Fe]  & rms  &  n& [Cu/Fe] &rms\\
\hline         
  986     &  5 & $-$0.10 & 0.10 &  6 & $-$0.05 & 0.08  &  5 & $-$0.03 & 0.14  &	  &	    &	   &   1 & $-$0.12 &	   &   12 & $-$0.07 & 0.11 &   &    &	\\    
  995     &  2 & $-$0.10 & 0.04 &  3 & $-$0.07 & 0.10  &  1 & $-$0.11 &	   &	  &	    &	   &	 &	   &	   &	3 & $-$0.16 & 0.09 &   &    &	\\    
\hline																					      
\end{tabular}																				      
\label{t:fegroup36}																			      
\end{table*}																				      
																					      
\clearpage																				      
\setcounter{table}{6}																			      
\begin{table*}																				      
\centering																				      
\caption[]{Abundances of $n-$capture elements in stars of NGC~362 with  												      
UVES spectra; n is the number of lines used in the analysis.}														      
\scriptsize																				      
\setlength{\tabcolsep}{1.3mm}
\begin{tabular}{rrccrccrccrccrccrccrcc}
\hline
  star    & n   &[Y/Fe]~{\sc ii}&rms&n& [Zr/Fe]~{\sc ii} & rms & n &[La/Fe]~{\sc ii}&rms&n& [Ce/Fe]~{\sc ii} & rms & n &[Nd/Fe]~{\sc ii} & rms   &  n  &[Eu/Fe]~{\sc ii}  & rms  &  n& [Dy/Fe]~{\sc ii} &rms\\
\hline         
1037      &  1  &  +0.00 & 0.14 &  2 & +0.46 & 0.06  &  1  & +0.30 & 0.16  &  2  &   +0.13 & 0.12  &  9  &  +0.28 & 0.16 &  1  & +0.66 & 0.12 &  1  & +0.75 & 0.17  \\ 
12017     &  1  &$-$0.13 & 0.14 &  2 & +0.40 & 0.11  &	1  & +0.17 & 0.16  &  2  & $-$0.06 & 0.12  &  9  &  +0.20 & 0.16 &  1  & +0.57 & 0.12 &  1  & +0.58 & 0.17  \\ 
\hline
\end{tabular}
\label{t:neutron36}
\begin{list}{}{}
\item[-] For species with only one transition, the scatter is estimated by summing in quadrature
the best fit variation obtained by changing temperature, gravity, metal abundance and microturbulent
velocity of amounts corresponding to the errors in these parameters. To this, we added in quadrature
a fit error of 0.05 dex.
\end{list}
\end{table*}

\clearpage

\setcounter{table}{7}
\begin{table*}
\centering
\caption[]{Abundances of Ba~{\sc ii} in stars of NGC~362. 
n is the number of lines used in the analysis.}
\begin{tabular}{rrcc}
\hline
   star  &  n &  [Ba/Fe]{\sc ii} & rms  \\
\hline   
  986  & 1 &   +0.400 &    \\  
  995  & 1 &   +0.447 &    \\  
\hline
\end{tabular}
\label{t:zrba36}
\end{table*}

\end{document}